___

# A New Nonlinear speaker parameterization algorithm for speaker identification[1]

*Mohamed Chetouani, Marcos Faúndez-Zanuy(\*), Bruno Gas, Jean-Luc Zarader*

Laboratoire des Instruments et Systèmes d'Ile-De-France (LISIF)
Université Paris Pierre et Marie Curie, PARIS, FRANCE
(\*) Escola Universitària Politècnica de Mataró, BARCELONA, SPAIN
e-mail: mohamed.chetouani@lis.jussieu.fr, faundez@eupmt.es

## Abstract

In this paper we propose a new parameterization algorithm based on nonlinear prediction, which is an extension of the classical LPC parameters. The parameters performances are estimated by two different methods: the Arithmetic-Harmonic Sphericity (AHS) and the Auto-Regressive Vector Model (ARVM). Two different methods are proposed for the parameterization based on the Neural Predictive Coding (NPC): classical neural networks initialization and linear initialization. We applied these two parameters to speaker identification. The fist parameters obtained smaller rates. We show for the first parameters how they can be combined with the classical parameters (LPCC, MFCC, etc.) in order to improve the results of only one classical parameterization (MFCC provides 97.55% and MFCC+NPC 98.78%). For the linear initialization, we obtain 100% which is great improvement. This study opens a new way towards different parameterization schemes that offer better accuracy on speaker recognition tasks.

## 1. Introduction

A key issue for implementing an accurate speaker recognition system is the set of acoustic features extracted from the speech signal. This set is required to convey as much speaker-dependent information as possible. The standard methodology to extract these features from the signal follows two trends: features are extracted through a filter bank processing or through linear prediction coding (LPC). Both the methods are to some extent linear procedures and are based on the underlying assumption that acoustic characteristics of human speech are mainly due to the vocal tract resonances, which form the basic spectral structure of the speech signal. However, human speech is a nonlinear phenomenon, which involves nonlinear biomechanical, aerodynamic, acoustic, and physiological factors, and LPC-derived parameters can only offer a sub-optimal description of the speech dynamics [7]. Therefore, in the last years there has been a growing interest for nonlinear models applied to speaker recognition applications.

In this paper we propose a new parameterization algorithm described in section 2. Next, we present the database and the other parameterization methods. Then, we briefly introduce the methods used for speaker identification. Finally, we give some preliminary results.

## 2. Neural predictive Coding

The Neural Predictive Coding (NPC) [11] model is a non linear extension of the LPC speech coding. This model is based on neural predictor of the speech waveform (cf. Figure 1).

The nonlinear prediction can be done by several methods like the Volterra filters [5] or neural networks [6]. The major advantage of the Volterra filters is that, like in linear predictors, the least mean square error solution for the filter coefficients can be expressed analytically. The main drawback lies in the fact that the number of coefficients grows fast with the prediction window dimension. A same drawback occurs with predictive neural networks. In addition, the weights solution cannot be expressed analytically in a function of the least mean square error.

The NPC model has the major advantage to allow a nonlinear modelization with an arbitrary limited number of coding coefficients. The NPC is used as speech encoder but only the output layer weights are considered as coding vector or feature vector. For that, the learning phase is realized in two times. First, the parameterization phase consists in the learning of all the weights by the prediction error minimization:

$$Q = \sum_{k=1}^{K} (y_k - \hat{y}_k)^2 = \sum_{k=1}^{K} (y_k - F(\mathbf{y}_k))^2 \qquad (1)$$

With $\mathbf{y}_k = \begin{bmatrix} y_{k-1}, y_{k-2}, \cdots y_{k-L} \end{bmatrix}^T$

---
[1] This work has been supported by COST-277, FEDER & CICYT TIC-2003-08382-C05-02

With $y_k$ speech signal, $\hat{y}_k$ predicted speech signal and k samples index and K the number of samples.

F is a non linear function which is composed of two functions $G_\mathbf{w}$ (corresponding to the hidden layer) and $H_\mathbf{a}$ (corresponding to the output layer):

$$F_{\mathbf{w},\mathbf{a}} = H_\mathbf{a} \circ G_\mathbf{w}$$

with $\hat{y}_k = H_\mathbf{a}(z_k)$ and $z_k = G_\mathbf{w}(\mathbf{y}_k)$

Where **w** denotes the hidden layer weights vector and **a** the output layer weights vector.

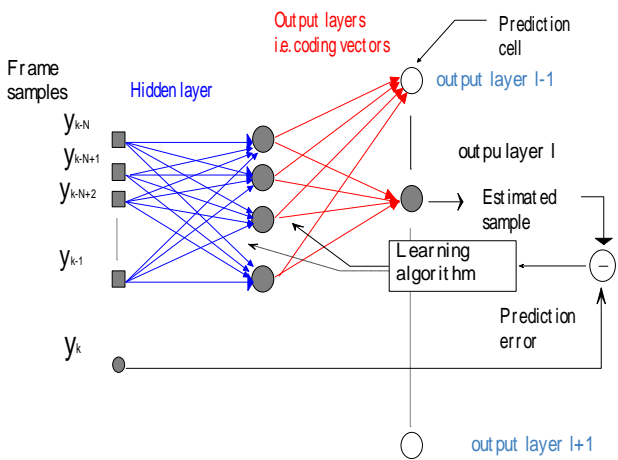

*Figure 1: The Neural Predictive Coding (NPC) architecture: a connectionist model used as a nonlinear predictor.*

In the parameterization phase, only the first layer weights **w**, which are the NPC encoder parameters are kept. Since the NPC encoder is set up by the parameters defined in the previous phase, the second phase, called the coding phase, consists in the computation of the output layer weights **a** : the coding vector.

F is a non linear function which is composed of two functions $G_\mathbf{w}$ (corresponding to the hidden layer) and $H_\mathbf{a}$ (corresponding to the output layer) :

$$F_{\mathbf{w},\mathbf{a}} = H_\mathbf{a} \circ G_\mathbf{w}$$

with $\hat{y}_k = H_\mathbf{a}(z_k)$ and $z_k = G_\mathbf{w}(\mathbf{y}_k)$

Where **w** denotes the hidden layer weights vector and **a** the output layer weights vector.

The NPC model was originally designed for speech feature extraction for phoneme recognition tasks [4]. Here, we present a new application for it, in feature extraction for speaker identification.

### 2.1. Coding phase

The coding phase consists in the estimation of the output layer weights by the prediction error minimization. Traditionally, these weights are initialized randomly. Here, we propose a new initialization method.

A good initialization for neural networks is very important. Currently, a good choice is to initialize the weights in a function of the training data [18]. Here, we propose to initialize the weights by considering the NPC as a linear model, which is in speech the LPC model.

### 2.2. Feature extraction for speaker identification

Currently, in speaker recognition task, the speech feature extraction is carried out in a same way for all the speakers. The key idea is to extract speaker-dependent characteristics by the NPC model (cf. Figure 2). Each NPC model is specialized in the processing of only one speaker. Contrary to other methods, here a speaker model is composed by a *feature extractor* and a *reference model*.

The speaker recognition task is done on two phases: the enrollment phase and the test phase. The NPC model is parameterized during the enrollment phase and it is used as a speech encoder during the test phase.

A sentence of one minute is used for the enrollment phase. The NPC parameterization phase is done a part of this sentence (only 12 seconds). Once the NPC is parameterized, the whole sentence is coded. This procedure is repeated for all the speakers belonging of the database. As a result, feature extractor and a classifier modelize each speaker.

The reference models are estimated by two methods: the Arithmetic-Harmonic Sphericity (AHS) [2] and the Auto-Regressive Vector Model (ARVM) [3,14].

During the test phase, all the NPC models code the testing speech and the identification is carried out.

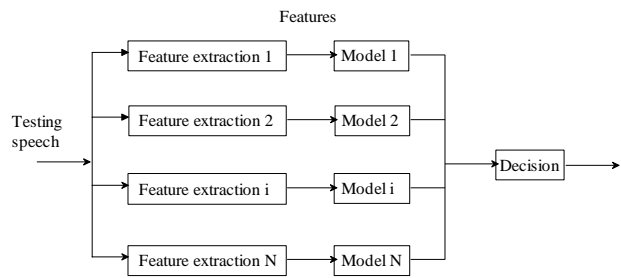

*Figure 2: Feature extraction architecture: the features are speaker dependent.*

## 3. Results

This section summarizes the experimental results.

### 3.1. Database

Our experiments have been computed over 49 speakers from the Gaudi database [1] that has been obtained with a microphone connected to a PC. The speech signal has been down-sampled to 8 kHz, pre-emphasized by a first order filter whose transfer function is $H(z)=1-0.95z^{-1}$. A 30 ms Hamming window is used, and the overlapping between adjacent frames is 2/3. A parameterized vector of order 16 was computed. One minute of read text is used for training, and 5 sentences for testing (each sentence is about 2-3 seconds long).

### 3.2. Parameterizations

We compare the proposed parameterization based on Neural Predictive Coding (NPC) with the following classical ones:

*3.2.1.  LPC*

The Linear Predictive Coding was a very commonly used method used in speech processing: recognition, synthesis, transmission, etc. ... This method is based on a linear modelization of the vocal tract. The all-pole auto-regressive (AR) coefficients model of the spectrum captures the vocal tract properties. Indeed, this model is more adapted for voiced sounds.

*3.2.2.  LPCC*

The Linear Predictive Cepstral Coding computes a LPC spectral envelope, before converting into cepstral coefficients. The LPCC are LP-derived cepstral coefficients. The LPCC is the most used coding method in speaker recognition.

*3.2.3.  MFCC*

The Mel Frequency Cepstral Coding is the most used speech coding method in recognition systems. The MFCC is based on signal decomposition with the help of filter bank, which uses the Mel scale. The MFCC results of a discrete cosine transform of the real logarithm of the short-term energy expressed on a Mel-frequency scale. The MFCC has shown good performances in speech recognition [12] but also in speaker recognition [15].

*3.2.4.  PLP*

The Perceptual Linear Predictive (PLP) [13] coding method is an example of knowledge integration resulting from psychoacoustics in the estimation of auto-regressive (AR) models. Indeed, this method integrates critical bands, equal loudness pre-emphasis and intensity-to-loudness compression. The PLP is based on the nonlinear Bark scale. It was originally designed to speech recognition with the removing of speaker dependent characteristics. However, it has been applied to speaker recognition and it has given good results [17].

### 3.3. Identification algorithms

In order to measure the performances of each parameterization method, we used two different methods: the Arithmetic-Harmonic Sphericity (AHS) [2] and the Auto-Regressive Vector Model (ARVM) [3]. The AHS is a statistical method based on second order measures while the ARVM is a predictive based method.

*3.3.1.  The Arithmetic-Harmonic Sphericity (AHS)*

A covariance matrix (CM) is computed for each speaker, and an Arithmetic-Harmonic Sphericity (AHS) measure is used in order to compare matrices [2]:

$$\mu(C_j C_{test}) = \log\left(tr(C_{test}C_j^{-1})tr(C_j C_{test}^{-1})\right) - 2\log(P)$$

Where $tr$ is the trace of the matrix, $P$ is the feature vector dimension, and the number of parameters for each speaker model is $\dfrac{P^2+P}{2}$ (the covariance matrix is symmetric).

For the CM model, more parameters imply a higher dimensional feature vectors.

*3.3.2.  The Auto-Regressive Vector Models (ARVM)*

The Auto-Regressive Vector Models (ARVM) [3,14] is a vectorial predictive method.  The speaker models are linear auto-regressive models, it is based on prediction of the q past parameters vector $\{\mathbf{x}_{t-1}, \mathbf{x}_{t-2},\ldots,\mathbf{x}_{t-q}\}$ :

$$\hat{\mathbf{y}}_t = \sum_{i=0}^{q} \mathbf{A}_i \cdot \mathbf{x}_{t-i} + e_t$$

where q is the model order, $\{\mathbf{A}_i\}$ are $P \times P$ matrices. $e_t$ is a vectorial white noise.

The matrices $\{\mathbf{A}_i\}$ are estimated by the help of the Levinson-Whittle-Robinson algorithm.

The identification process is carried out by the help of a symmetric distance.

### 3.4. Identification results

*3.4.1.  Identification by the AHS method*

Table 1 shows the identification rates for different parameterization schemes. One can see that, for the traditional, the best performances are obtained for the MFCC (97.55%) and the LPCC (96.73%) coding methods, which it is in agreement with the coding characteristics. Indeed, these methods try to model the phonetic context but also the speaker characteristics. The LPC model has a better score (90.61%) than the PLP (86.12%). This is due to the fact that the PLP method suppresses speaker dependent characteristics. It is why the PLP allows comparable performances with the MFCC in speech recognition task.

| PARAMETERIZATION | IDENTIFICATION RATE (%) |
|---|---|
| LPC | 90.61 |
| LPCC | 96.73 |
| MFCC | 97.55 |
| PLP | 86.12 |
| NPC (random initialization) | 61.63 |
| NPC (linear initialization) | 100 |

Table 1. Experimental results for different parameterizations with the AHS method.

Although it is evident that the random initialization for the NPC model is far from the state-of-the-art parameterizations, we think that this is a first step towards the proposal of better discriminative nonlinear features. On the other hand, this proposal differs from [8,9] that proposed the residual signal obtained by means of a nonlinear filtering as a distance measure. Thus, this paper presents an approach more similar to the classical parameterization schemes. However for the linear initialization, one can see that the performances are better (100%) than the state-of-art parameterizations.

In order to measure the performances of the NPC model, we use another method based: the ARVM.

### 3.4.2. *Identification by the ARVM method*

Table 2 shows the identification rates for different parameterization schemes. The results for all the parameters are practically identical. But, one can see that the NPC performances are improved by this method. One of the reason for this difference is that the AHS method is a unimodal method which is restricted to model linear correlations [17]. Moreover, the NPC method is nonlinear consequently the behavior is different.

| PARAMETERIZATION | IDENTIFICATION RATE (%) |
|---|---|
| LPC | 90.61 |
| LPCC | 93.06 |
| MFCC | 95.69 |
| PLP | 78.36 |
| NPC (random initialization) | 88.57 |
| NPC (linear initialization) | 100 |

Table 2. Experimental results for different parameterizations with the ARVM method.

One can also notice that the linear initialization allows obtaining the best results (100%).

### 3.4.3. *Identification by fusion in the AHS method*

In order to improve the AHS results for the random initialization, we have evaluated the combination between classical parameterizations and the proposed new scheme, in a similar way that [8,9]. This kind of combination is also known as opinion fusion [10].
For this purpose, we have done the following steps:

1. Distance normalization [16] $o_i' = \dfrac{1}{1+e^{-k_i}}$, where

   $k_i = \dfrac{o_i - (m_i - 2\sigma_i)}{2\sigma_i}$, $o_i$ is the opinion of

   classifier $i$, $o_i' \in [0,1]$ is the normalized opinion, $m_i, \sigma_i$ are the mean and standard deviation of the opinions of classifier $i$ using the genuine speakers (intradistances).

2. Weighted sum combination with trained rule $O = \alpha o_1 + (1-\alpha) o_2$, where $o_1$, $o_2$ are the scores (distances) provided by each classifier, and α is a weighting factor.

Table 3 shows the experimental results for several parameterizations' combinations.

|  | LPCC | MFCC | PLP | NPC |
|---|---|---|---|---|
| LPC | 97.14 | 99.18 | 94.69 | 93.47 |
| LPCC |  | 98.78 | 97.96 | 97.55 |
| MFCC |  |  | 97.55 | 98.78 |
| PLP |  |  |  | 91.84 |

Table 3. Identification rates for several combinations

Table 1 reveals that although the NPC features alone do not perform good enough, their combination with the classical ones outperform the identification rates of LPC, LPCC, MFCC and PLP. Table 4 shows the combination factor α.

|  | LPCC | MFCC | PLP | NPC |
|---|---|---|---|---|
| LPC | 0.59 | 0.63 | 0.51 | 0.3 |
| LPCC |  | 0.38 | 0.27 | 0.15 |
| MFCC |  |  | 0 | 0.22 |
| PLP |  |  |  | 0.23 |

Table 4. Selected combination factor α for the results shown in table 3.

## 4. CONCLUSIONS

**Final paper**
We are trimming our nonlinear parameterization in order to improve the speaker recognition rates using only this parameter. The NPC with a linear initialization gives the best results (100%). Anyway, the combination of classical (LPCC, MFCC) and the NPC parameters improve the results and offer a new approach to the speaker recognition task. Moreover, we are investigating the NPC behavior and models computation (AHS, ARVM) in speaker recognition.